\RequirePackage{fix-cm}

\documentclass[a4paper,11pt]{article}

\usepackage{graphicx}
\usepackage[utf8x]{inputenc}
\usepackage{todonotes}
\usepackage{subfigure}
\usepackage{hyperref}
\usepackage{pgfplots}

\pgfplotsset{table/col sep=semicolon}

\title{Improving the Scalability of DPWS-Based Networked Infrastructures}
\author{Filipe Campos, José Pereira\\
\begin{small}High-Assurance Software Laboratory\end{small}\\
\begin{small}INESC TEC \& University of Minho\end{small}\\
\begin{small}\{fcampos, jop\}@di.uminho.pt\end{small}\\\\
Technical Report\\\\
}

\date{\today}

\begin{document}

\maketitle

\begin{abstract}
The Devices Profile for Web Services (DPWS) specification enables seamless discovery, configuration, and interoperability of networked devices in various settings, ranging from home automation and multimedia to manufacturing equipment and data centers. Unfortunately, the sheer simplicity of event notification mechanisms that makes it fit for resource-constrained devices, makes it hard to scale to large infrastructures with more stringent dependability requirements, ironically, where self-configuration would be most useful.

In this report, we address this challenge with a proposal to integrate gossip-based dissemination in DPWS, thus maintaining compatibility with original assumptions of the specification, and avoiding a centralized configuration server or custom black-box middleware components. In detail, we show how our approach provides an evolutionary and non-intrusive solution to the scalability limitations of DPWS and experimentally evaluate it with an implementation based on the the Web Services for Devices (WS4D) Java Multi Edition DPWS Stack (JMEDS).
\end{abstract}

\section{Introduction}
\label{sec:intro}
Service Oriented Architectures (SOA) are a mainstay of enterprise computing and there is now a growing interest in services for systems of connected devices in a variety of environments, ranging from industrial manufacturing equipment to home automation. The first enabler for this is that it has become cost-effective to equip most devices with a considerable a{\-}mount of processing and networking resources, making Ethernet networking and mainstream operating systems (e.g. Linux) ubiquitous. The second enabler is the observation that SOA provides the best solution for interoperability, composability, and long term maintainability, challenges that are particularly acute in these scenarios. In fact, the current trend in connected devices is expected to accelerate as the vision for the Internet of Things becomes a reality.

In this context, being able to expose the joint capabilities of large sets of devices as logical services is enticing, as much as coordinating and composing services in business processes has been important in enterprise service-oriented computing. Besides the obvious issue of efficiently routing information to all destinations, scalable and dependable dissemination raises a number of traditionally hard problems. For instance, the effort involved in keeping track of destinations in a dynamic environment, namely, how the effort is spread across different components\,\cite{Stoica:2001p2290}. Or how reliable delivery (end-to-end or multi-party) is ensured, namely, how to manage acknowledgement and retransmission without depending on any single system component\,\cite{Bir99}. Or even, how to pace transmission in order not to overwhelm any particular destination\,\cite{614104}.

Two main approaches have been proposed to deal with the need for dependable and efficient information dissemination in service-oriented computing.
The first approach is to embody the information dissemination logic in middleware, namely, by using an Enterprise Service Bus or by using Java Messaging Service (JMS) or Extensible Messaging and Presence Protocol (XMPP) transports for Web Services\,\cite{jms_transport,xmpp_transport}. This approach takes advantage of mature, tried and tested technology which solves the aforementioned challenges and is already deployed and well-known in many organizations. On the other hand, this approach introduces a dependency on a specific software stack and often, on centralized, albeit redundant through replication, for instance, messaging servers. Moreover, this middleware is often not available for the entire range of devices that need to be supported, either because it assumes computing resources of typical enterprise systems or simply because the middleware vendor does not target the desired platform. An alternative to the use of such a middleware platform would be to introduce `superpeers' in the system. Finally, by providing messaging as a black box, it supports only a fixed range of information exchange patterns and limits the range of problems and environments that can be addressed.
The second approach has been to provide information dissemination at the service level, by means of specifications that can be combined to expose different message exchange patterns with various levels of QoS. An example of this approach is the WS-Notification\,\cite{ws_bn,ws_brn,ws_t} family of standards, which besides simple notification and subscription, supports topic based publish-subscribe and brokered dissemination for scalability. An alternative is the simpler WS-Eventing specification\,\cite{ws_eventing} which, although lacking explicit support for brokered dissemination, embodies a flexible filtering mechanism in the base specification and favors lightweight implementations and the many-to-one dissemination scenario. It has therefore been the preferred choice for connected devices, namely, within standards like WS-Management\,\cite{ws_man} and Devices Profile for Web Services (DPWS)\,\cite{dpws}. Both eventing specifications can be combined with other standards, such as WS-ReliableMessaging (WS-RM)\,\cite{wsrm} for end-to-end acknowledged message delivery or WS-AtomicTransaction (WS-AT)\,\cite{wsat} for multi-party transactional atomicity guarantees.

Note, however, that both approaches, at the middleware and service levels, emphasize one-to-many communication and do it by means of setting up a centralized broker infrastructure. Dependability rests on the central server supporting replication in a cluster, while end-to-end reliability and atomicity depend on participants having sufficient memory for buffering and stable storage for a transactional log. In short, the assumption of a centralized heavyweight infrastructure permeates most existing solutions for information dissemination in service-oriented computing.

DPWS defines a set of protocols that resource constrained devices should implement in order to achieve seamless networking and interoperability through Web Services. It assumes that each device behaves as a standard \emph{hosting service}, providing basal functionality, and exposing one or more \emph{hosted services} that offer device specific functionality. Besides basic SOAP, WSDL, the HTTP binding, WS-Ad\-dress\-ing, and WS-Security that are at the core of Web Services capabilities and interoperability, DPWS also includes WS-E\-vent\-ing, as previously mentioned, SOAP-over-UDP\,\cite{soap_over_udp}, which enables the usage of UDP as a transport for SOAP messages and enables network level multicast, thus paving the way for dynamic discovery, enabled by combining WS-Discovery\,\cite{ws_disc}, WS-Meta\-data\-Exchange\,\cite{ws_me}, and WS-Policy\,\cite{ws_p}. These protocols allow a client to discover devices in the network, and to learn about their services, resources, characteristics.
Multiple implementations of the DPWS exist\,\cite{ws4d_url,soa4d_url}, and modern operating systems, such as Windows Vista, Windows Embedded CE, and Windows 7 are shipped with a built-in DPWS framework, thus rendering this specification available in most personal computers and in many devices such as set-top boxes.

Although DPWS provides an adequate infrastructure for small scale systems, such as home automation, it is becoming increasingly interesting when managing large number of components, albeit its scale limitations. First, the use of WS-\-E\-vent\-ing imposes a burden on the publisher, that has to notify all subscribers. Moreover, when a resource exposed by many devices has to be updated, e.g. to change a configuration variable, the initiator device must contact every destination individually. Finally, as there is no support for transactional coordination mechanisms, such lengthy operations involving large numbers of destinations are susceptible to faults and cannot be restarted or recovered if stopped. This is particularly worrisome as such notifications and configuration updates may correspond to critical alerts and urgent commands.
It does not make sense to resort to heavyweight coordination protocols such as WS-Coordination\,\cite{wscoor} and WS-AT in such a scenario, because, even if devices could support their requirements, they would not scale to hundreds of participants. Thus, a scalable lightweight coordination protocol, that fits the general DPWS assumptions, is necessary.

In this report, we address this need by introducing a service oriented architecture for information dissemination based on existing standards and distributed gossiping.
Gossiping is a lightweight approach for information dissemination that has inherent scalability and atomicity guarantees, while being simple, resilient, and frugal on resources.
Although one could easily use an existing gossip-based messaging middleware as a black box transport protocol (e.g. NeEM\footnote{http://neem.sf.net}), this would still have many of the shortcomings of the first approach outlined above.
Instead, our Web Services Gossip (WS-Gossip) framework provides operations that can be combined to architect a variety of gossip-style interactions, such as the \textit{push} vs. \textit{pull}, \textit{eager} vs. \textit{lazy}, and \textit{infect-and-die} vs. \textit{balls-and-bins} gossip variants, to address multiple applications and environments, and furthermore, can be integrated with different membership management strategies, according to the scale and dynamics of each system.
WS-Gossip also enables the usage of gossip by participants and their clients, while at the same time minimizing the impact on existing producers and consumers due to its adaptability.

In comparison to WS-PushGossip\,\cite{1462812} and WS-Membership\,\cite{Vogels:2003p1034}, which are both based in the WS-Coordination standard, the proposed framework is better suited for devices and their inherent limitations, as it leverages standards which are widely available in devices, and as the implementation and use of such a resource consuming protocol is not necessary or might not even be feasible. Moreover, WS-PushGossip only provides \textit{push} dissemination, while the proposed framework can provide four different variants of gossip dissemination, as well as gossip-based aggregation. And, WS-Membership implies that specific monitoring agents are capable of assessing the availability of services with specific mechanisms, whereas the proposed framework relies on WS-Discovery, which is mandatorily available in DPWS compatible devices, for the same purpose.
We illustrate and evaluate our approach by implementing and applying it in the context of the DPWS, more concretely, by comparing it with WS-Eventing.

The remaining of this report is structured as follows. Section~\ref{sec:bg} provides background information by briefly describing various Web Services standards. Section~\ref{sec:protocol} houses the main contribution, by casting gossip functionality as a set of services that can be combined with existing standards. Then, Section~\ref{sec:impl} experimentally evaluates the performance of the implementation of WS-Gossip on the Web Services for Devices (WS4D)\,\cite{ws4d_url} Java Multi Edition DPWS Stack (JMEDS).
Section~\ref{sec:related} presents other related works, comparing them to the current approach.
Section~\ref{sec:concl} concludes the paper with some remarks on the presented framework and results, and pointing possible directions for future work.

\section{Background}
\label{sec:bg}
Coordination of services involving information dissemination are concerned with
two primary challenges:
\begin{itemize}
 \item How can information be efficiently disseminated to all relevant targets?
This includes the propagation of information itself but also the maintenance of
target lists.
 \item What guarantees are given that information is actually disseminated to
all relevant targets? This is particularly relevant when there are failed nodes
and network faults.
\end{itemize}

The following sections describe how existing Web services standards address
these concerns and how gossip-based protocols provide an interesting solution to
both of them.

\subsection{Service Oriented Standards}

\subsubsection{Eventing and messaging}

The WS-Eventing specification supports simple
publisher-subscriber interaction, by defining how Web Services can subscribe to
or accept subscriptions for event notification messages\,\cite{ws_eventing}.

The specification defines the four following roles:
\begin{description}
 \item [Event Source]Sends notifications on triggered events,
and accepts requests for creating subscriptions.
 \item [Subscription Manager]Manages event subscriptions. 
It notifies \textbf{Subscribers} when their subscriptions are terminated unexpectedly,
 and replies to their subscription management enquiries, such as subscription's
status retrieval, renewal or deletion.
 \item [Event Sink]Receives event notification messages.
 \item [Subscriber]Contacts an \textbf{Event Source} to create a subscription to
manifest the interest of its associated \textbf{Event Sink} to be notified on the
occurrence of some event.
notifications. It is also responsible for issuing subscription management
requests to the \textbf{Subscription Manager}.
\end{description}

In the simplest scenario, with only two intervening Web Services, a \textbf{Publisher}
will comprise both the \textbf{Event Source} and the \textbf{Subscription 
Manager} roles, whereas the entity that accumulates both the
\textbf{Subscriber} and \textbf{Event Sink} roles will be referred as a \textbf{Subscriber}.

\begin{figure}[htbp]
 \centering
\includegraphics[width=.8\textwidth]{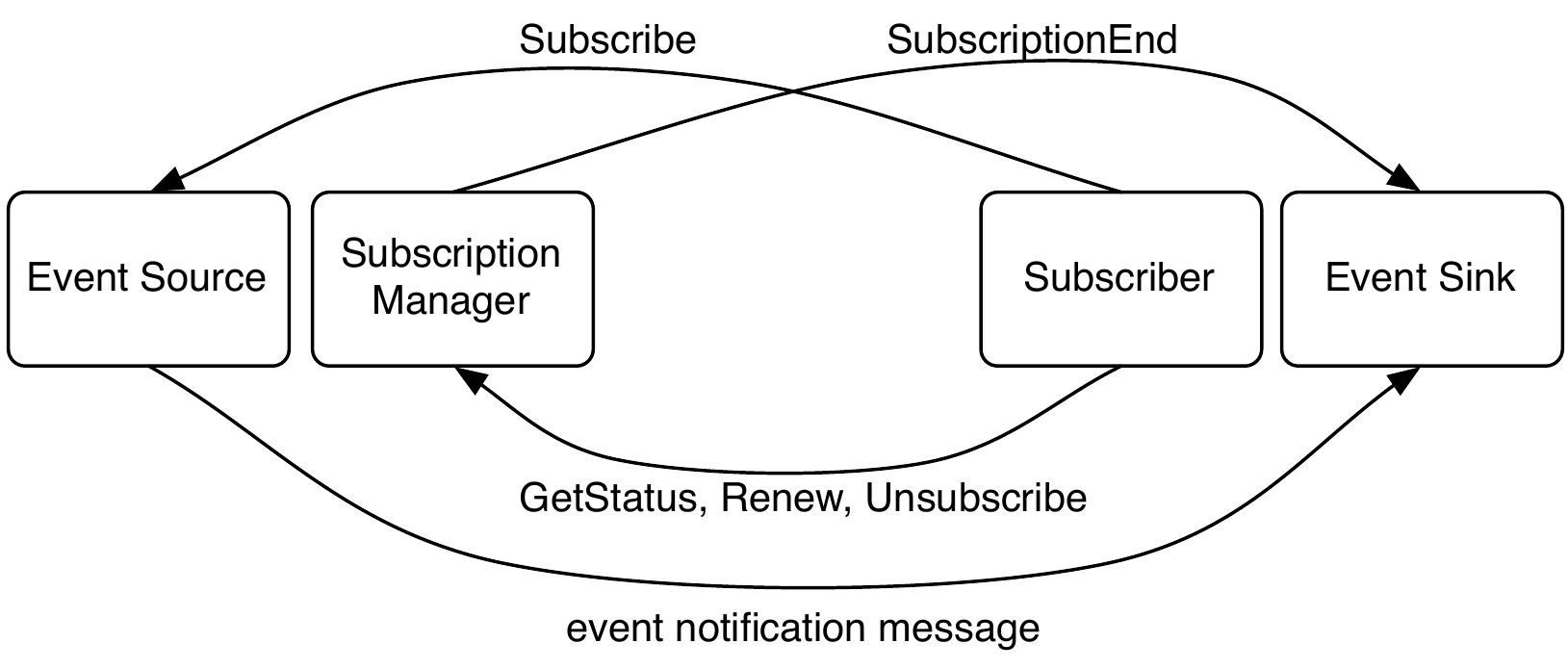}
\caption{WS-Eventing components.}
\label{fig:ws_eventing}
\end{figure}

WS-Eventing defines the concept of delivery mode, in order to better adapt the
general publish/subscribe pattern to scenarios with different event delivery
requirements. The default delivery mode of this specification is the single
delivery asynchronous \textit{push} mode. But, for instance, situations with slow event
consumers where they poll for event messages may be preferred, in order to
control the rate of message arrival and avoid overwhelming them if the rate of
generation and transmission of event messages is far superior to their
processing rate.
 The subscription request message defines the specific delivery mode to be used in
notifying the identified \textbf{Event Sink}, and new delivery modes can be freely used,
if both event sources and consumers support them. In the event that a \textbf{Subscriber}
requests a delivery mode that is not supported by the \textbf{Event Source}, it will
respond signaling this situation, and it may convey a list of the supported
delivery modes.
This specification proposes, as an example, that notifications can be wrapped in
a standard message instead of the default unwrapped mode where each notification
is transmitted as a message typed according to the event's action.

Although it lacks explicit support for brokered dissemination, it embodies a
flexible filtering mechanism in the base specification, favoring lightweight
implementations and many-to-one dissemination scenarios. And since it was backed
by major vendors, such as IBM, Microsoft or TIBCO, it has therefore been the
preferred choice for connected devices, namely, within WS-Management\,\cite{ws_man} and DPWS.

The alternative family of standards OASIS WS-Notification\,\cite{ws_bn,ws_brn,ws_t} also provides, besides simple notification and subscription mechanisms, extensible topic definition and brokered dissemination.

Since the early 1990s, Reliable Messaging has been seen as a solution for such
scenarios by the IT community, and so, several message queueing technologies
have been used, such as IBM's WebSphereMQ and Microsoft's MSMQ, in addition to
reliable publish/subscribe technologies, such as Tibco Rendezvous. In an effort
to bridge all these different technologies, the Java Message Service (JMS) API
was developed by the Java Community Process. Some of these technologies were
adapted to Web Services. However, due to the exploitation of proprietary
protocols, interoperability can only be achieved recurring to gateways that
translate between specific pairs of environments.

With the emergence of Web Services as the preferred integration solution for
distributed systems, WS-Reliable\-Mes\-sag\-ing (WS-RM)\,\cite{wsrm} is the
currently adopted standard for achieving reliable message exchange between
distributed applications in the presence of software component, system and
network failures. Due to the interoperable nature of Web Services, WS-RM allows
to bridge two different infrastructures, such as different operating or
middleware systems, into an end-to-end model where messages are exchanged
reliably\,\cite{ibm_secure}. So, this standard ensures the interoperability of
services in what comes to
Reliable Messaging, which also simplifies the development of services,
since they must implement the protocols, minimizing the number of errors in
business logic\,\cite{ibm_secure}.

\begin{figure}[htbp]
\centering
\includegraphics[width=.6\textwidth]{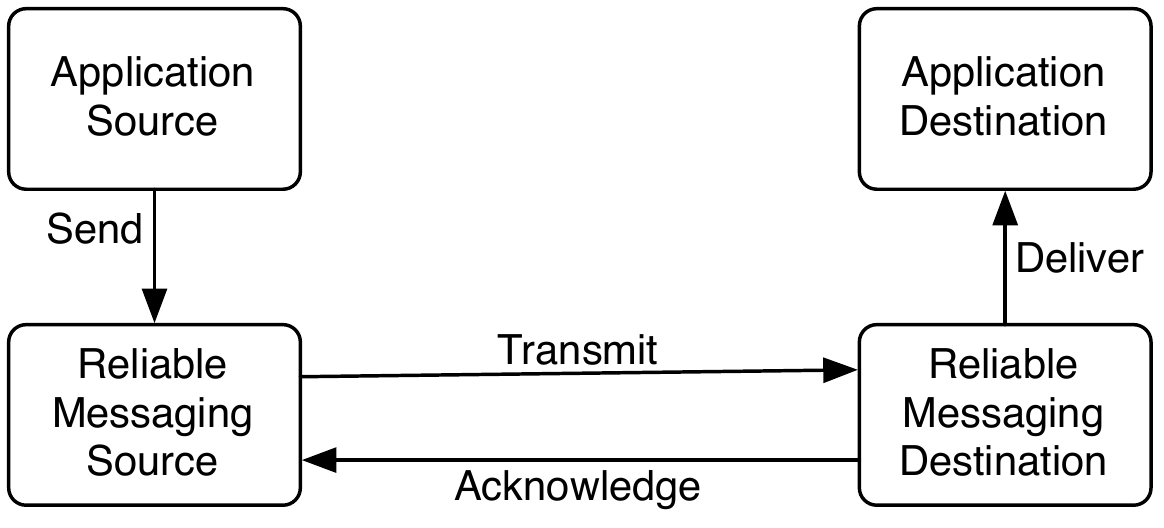}
\caption{WS-ReliableMessaging basic interaction.}
\label{rm_fig}
\end{figure}

WS-RM distinguishes all the entities
involved in an interaction, as well as the various meanings of the terms
\textit{send}, \textit{transmit}, \textit{receive} and \textit{deliver}, as they
relate to different components. In that sense, the basic model of
WS-RM is described in Figure\,\ref{rm_fig} and it
includes four distinct entities:
\begin{description}
\item [Application Source]Service or application logic that \textit{sends}
the message to the \textbf{RM Source};
\item [RM Source]Physical processor or node that performs the actual wire
transmission;
\item [RM Destination]Target processor or node that \textit{receives} the
message and then \textit{delivers} it to the \textit{application destination};
\item [Application Destination]Target service of the message.
\end{description}

These nodes are endpoints, which according to the WS-RM
standard, represent addressable entities that send and receive Web Services
messages.

The basic mechanism of the standard works, in a simplified way, as follows: the source
node sends a Web Service message containing a WS-RM
header, which is received by the destination node that then replies by sending
an acknowledgment message to the source node.

There are several types of assurances defined in WS-RM, in terms of message
delivery:
\begin{description}
\item [AtMostOnce]A message is delivered at most once, but it might not be
delivered at all.
\item [AtLeastOnce]A message is delivered at least once, but it could be
delivered more times.
\item [ExactlyOnce]This type is a combination of the previous two. A message is
delivered only once.
\item [InOrder]When there are several ordered messages, they are delivered in
the same order as they were sent.
\end{description}
Each reliable message exchanging sequence will enforce the message delivery guarantees according to the type defined in the sequence's creation.

\subsubsection{Transactional Coordination}

The term coordination sometimes refers to a type of orchestration that is
defined in the WS-Coordination specification\,\cite{wscoor}. It
specifies an extensible framework for \textit{context} management, that provides
coordination for the actions of distributed applications. This coordination is
achieved through provided protocols that support distributed applications, for
instance, those that need to reach consistent agreement on the outcome of
distributed transactions.

An application service can create a context needed to propagate coordination
information to other services involved in some activity. These services will
then need to register as participants of that activity. For this purpose, the
application must include the created \textit{coordination context} in the
messages that it sends to the referred services.

A \textit{coordination context} can be transmitted using application-specific
mechanisms, such as the header element of a SOAP application message. This kind
of conveyance is commonly referred to as flowing the \textit{context}.

The structure of a \textit{context} and the requirements to propagate it between
cooperating services are also defined in WS-Coordination, and can depend on the
type of coordination that is used. A \textit{coordination context} contains
information on:
\begin{itemize}
\item how to access a coordination registration service;
\item the coordination type;
\item relevant extensions.
\end{itemize}

This framework also enables existing transaction processing, workflow, and other
systems for coordination to hide their proprietary protocols and to operate in
an heterogeneous environment.

This specification is not sufficient by itself to coordinate Web Services, since
it provides only a coordination framework, leaving undefined the concrete
protocol and targeted coordination type. The standards WS-Atomic{\-}Transaction (WS-AT)\,\cite{wsat} and
WS-BusinessActivity (WS-BA)\,\cite{wsba} implement the WS-Coordination framework and also extend it 
by defining their own coordination type: short-term atomic transactions, and
long-running business activities, respectively.

The WS-AT specification defines a
protocol that can be plugged into WS-Coordination to provide an adaptation for
Web Services of the classic \textit{2PC}\,\cite{understanding_soa} mechanism, making the changes,
resulting from the activity of some service, persistent. It
is often said that this protocol does not adapt well to Web Services.
Nonetheless, it is adequate for interoperability across short-lived, co-located
services that need to ensure consistent, all-or-nothing results for a
transaction.

A \textit{2PC} process
consists on a poll conducted by the \textit{coordinator} that will lead it to
send two alternative directives to all the participants in the transaction:
	\begin{description}
	 \item [commit]If all of the registered services have responded
indicating that the changes were successful.
	 \item [rollback]If at least one of the registered services
fails to respond or responds indicating a failure. 
	\end{description}

A service that participates in an atomic transaction can register for more than
one of the different types of coordination protocols, as defined in the WS-AT
specification:

	\begin{description}
	\item [Two-Phase Commit]Coordinates registered participants to
reach a commit or abort decision, and ensures that all participants are informed
of the final result. It has two variants:
		\begin{description}
		\item [Volatile 2PC]Participants manage volatile
resources such as a cache register or a window manager.
		 \item [Durable 2PC]Participants manage durable
resources such as a da{\-}ta{\-}base register or a file.
		\end{description}
	 \item [Completion]Initiates commit processing when an
application tells the \textit{coordinator} to either try to commit or abort an
atomic transaction. Based on the registered participants for each protocol, the
\textit{coordinator} begins with \textit{Volatile 2PC} and then proceeds through
\textit{Durable 2PC}. After the transaction has completed, a status is returned
to the application and the final result to the service that initiates the
transaction (\textit{initiator}), if it has registered for this protocol.
	\end{description}

\subsubsection{Combining services}

One of the main benefits of using standards lies in the ability to combine them due
to their well known interfaces and behavior, in order to extract the most suitable features for each scenario.

For instance, both of the aforementioned event-driven specifications, WS-Eventing and WS-Notification, can
be combined with a coordination protocol in
order to guarantee the atomicity of an event\,\cite{4279671}. Although this
composition ensures that notifications reach all the relevant targets, it proves
to be a heavy process for resource constrained devices, specially in scenarios
with a large number of targets, or with a large amount of communication errors,
where WS-RM could help mitigating them, but also at the cost of increasing
resource consumption.

WS-RM can be combined with several WS-* standards. On the one hand, it can
improve its features by leveraging:
\begin{itemize}
 \item WS-Addressing, which enables the identification of messages and addresses
of
endpoints. This specification was modified to accommodate some needs of the WS-RM
specification, like the reuse of a message ID when retransmitting identical
messages to counter communication errors\,\cite{soa_concepts_tech_design};
 \item WS-Security, to protect the integrity and confidentiality of the
exchanged messages;
 \item WS-Policy, to specify the delivery assurance, among other requirements,
for a
sequence\,\cite{soa_concepts_tech_design,soa_field_guide,soa_approach}.
\end{itemize}
On the other hand, due to its ability to ensure reliable communication between
two endpoints, WS-RM can be leveraged by other standards, such as
WS-Eventing, WS-Notification, WS-AT, WS-BA and WS-Coordination to
achieve reliable communication among the intervening parties.

Although the WS-RM specification allows to condition
service activities, it is different from WS-AT or
WS-BA, in the sense that a coordinating entity is not
needed to inspect the progress of the activities, being the reliability rules
conveyed as SOAP headers in the exchanged
messages\,\cite{soa_concepts_tech_design}.

Regarding the questions posed in the beginning of this section, WS-RM would be a 
suitable standard to ensure point-to-point reliable message delivery. However, it
would be very inefficient and poise a heavy weight on the message sender in terms 
of processing power, if there are lots of message recipients or if lots of errors 
occur. In order for WS-RM to guarantee atomic delivery to all targets, it would have to 
rely on WS-AT, or a similar protocol based in WS-Coordination, which would, once again, increase the
consumption of the sender's processing and communication resources, due to the additional message traffic.

\subsection{Gossip}

In computer networking, gossiping describes the process where a participant that
intends to disseminate some information chooses a small random subset of other
participants and forwards the information to them. Each of these destinations,
upon receiving the information, repeats the same procedure, hence, the gossip
moniker. This procedure mimics also how epidemics spread in populations and,
therefore, are also known as epidemic protocols\,\cite{Eugster:2004p10747}.

\subsubsection{Reliability and Scale}

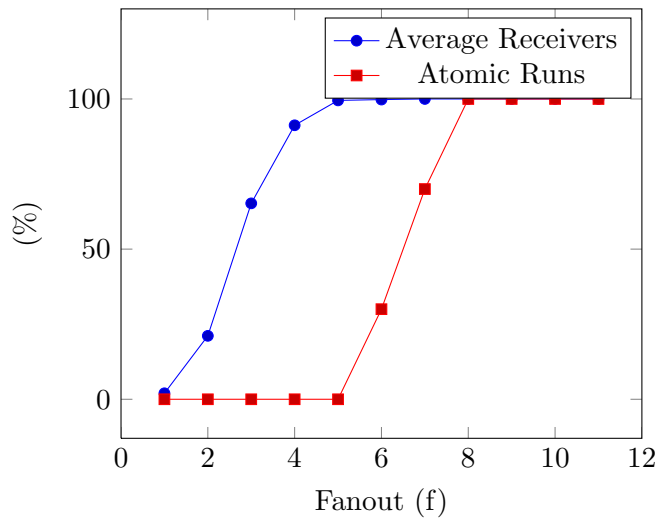
\begin{figure}[t]
 \centering
 \begin{tikzpicture}
  \begin{axis}[xlabel=Fanout (f),ylabel=(\%),ymax=130]
   \addplot table[header=false,x index=0,y index=1]{gossip.dat};
   \addplot table[header=false,x index=0,y index=2]{gossip.dat};
   \legend{Average Receivers,Atomic Runs}
  \end{axis}
 \end{tikzpicture}
\caption{Reliability of gossip (250 participants, 10 dissemination runs,
variable fanout).}
\label{fig:gosreliability}
\end{figure}

Most interestingly, gossip protocols don't need a reactive mechanism to deal
with failures, namely, buffering, acknowledgement, retransmission, and garbage
collection, which account for most of the complexity in common communication
protocols. Instead, reliability is proactively achieved by the protocol's
inherent redundancy and randomization, that cope with both process and network
link failures.

The expected probability for a message being delivered to each destination and
to all destinations as a whole can be derived directly from protocol parameters
$f$, the number of targets that are locally selected by each process for
gossiping, and $r$, maximum number of times a message is relayed before being
ignored. Figure~\ref{fig:gosreliability} illustrates the impact of these parameters
by showing simulation results of disseminating 10 messages to 250 receivers,
with $r=5$ and a variable $f$. Notice that with $f>4$ each destination gets each
message with a very high probability. With $f>7$, each message is atomically
received by all destinations also with a very high probability.

By adjusting $r$ and $f$ parameters according to system size and expected
faults, gossip can be configured such that any desired average number of
receivers successfully get the message. Better yet, parameters can be set such
that the message is atomically delivered to all the receivers with high
probability leading to guaranteed atomic delivery\,\cite{1297243}. The key to
scalability is that the required fanout configuration is at worst
logarithmically proportional to system size.

\subsubsection{Variants}

There are two main variants of gossip protocols\,\cite{892324}, which provide
different message exchange patterns and performance trade-offs. In \emph{push
gossip}, a node that becomes aware of new information, conveys it immediately to
target nodes. This variant is adequate for one-to-many dissemination of small
messages and events. With \emph{pull gossip}, instead of gossiping upon arrival
of new information, a node periodically selects a number of peers and asks them
for new information. It has been shown that combining \emph{push} and \emph{pull
gossip} results in dissemination being achieved in a lower number of
steps\,\cite{892324} and provides a generic framework for gossiping that can be
tailored for multiple purposes by parameterizing it with different aggregation
functions\,\cite{newscast03}.

In addition, lazily deferring the transmission of payload improves performance
in heterogeneous networks, allowing gossip protocols to approximate ideal
resource usage efficiency\,\cite{Pereira06}. Such \emph{lazy} variants are most
useful when the data payload is very large, but also when it is very likely that
the data is already known throughout the network.

Finally, there are two options regarding relaying duplicate messages. In the
\textit{infect-and-die} model, a participant that receives the message (i.e. is
infected), sends the received message to other nodes, and then never sends it
again, becoming dead in the analogy with epidemics. In the
\textit{infect{\--}forever} model, also known as
\textit{balls-and-bins}\,\cite{koldehofe02simple}, a participant might relay
received message multiple times, possibly until $r$ rounds are reached. This
last alternative has the advantage of requiring no state at participants to
recall recently relayed messages. On the other hand, it usually requires more
network resources as the relay limit has to be set conservatively.

\subsubsection{Membership Management}

A key component of a gossip protocol is the ability to obtain random subsets of
participants to direct messages at in each gossip operation. This component has
to provide an uniform random sample and, as much as possible, drawn from a
current view of operational participants\,\cite{1045666}. The first option is to
share the full list of participants, allowing each of them to locally draw
subsets as desired\,\cite{312207}. This is adequate when the list does not
change frequently, to avoid taxing the network with constant updates, and is
small enough to fit each participants memory.

If these conditions are not met, it has also been shown that sufficiently good
random samples can be obtained by having each participant keep a small partial
view of the system, which is itself maintained using a gossip
protocol\,\cite{945507,Eugster:2004p10747}. A particularly simple but effective
approach\,\cite{Voulgaris:2005p6235} is allowing a node to exchange some
elements in its local list with the same number of elements from some other
node. This progressively shuffles the list of each participant and leads to the
desired uniform random sample. By adding a time-based lease and renewal
mechanism, it also deals with participants entering and leaving the system.

\subsection{Motivation}
\label{sec:motiv}
Providing comprehensive support for gossip-based information dissemination in Web Services, in a way that integrates with existing enterprise information systems, requires however that one answers the following questions:
\begin{itemize}
 \item How to allow different gossip variants, namely, regarding \textit{push}\,vs.\textit{pull}, \textit{eager}\,vs.\textit{lazy}, and \textit{infect-and-die}\,vs.\textit{balls-and-bins} in a common framework, to address multiple applications and environments?
 \item How to support a range of membership management strategies, fit for different system scales and dynamics?
 \item How to enable usage of gossip in participants and their clients, while at the same time minimizing the impact on producers and consumers, namely, regarding required middleware?
\end{itemize}
In this section, we address these challenges with a set of specifications of service port types, SOAP headers, and policy assertions that can be used to compose a variety of solutions.

Regarding the target of this proposal, i.e. largely heterogeneous devices, we mainly envision the use of gossip to replace the existing mechanisms of alerts and events propagation for scenarios with a large number of targets.
Hence, we find gossip as an alternative to notification of events when:
\begin{itemize}
  \item a publisher must deal with many subscribers, which will consume a considerable amount of resources in subscriptions storage and maintenance;
  \item important messages or critical alerts must be conveyed;
  \item high rate of messages;
\end{itemize}
WS-Gossip could also serve as a complement to WS-Eventing, if devices are able to switch from one to the other according to the number of subscribers and the nature of the event.

\section{Gossip Service}
\label{sec:protocol}
\begin{figure}[htbp]
\centering
\includegraphics[width=.75\textwidth]{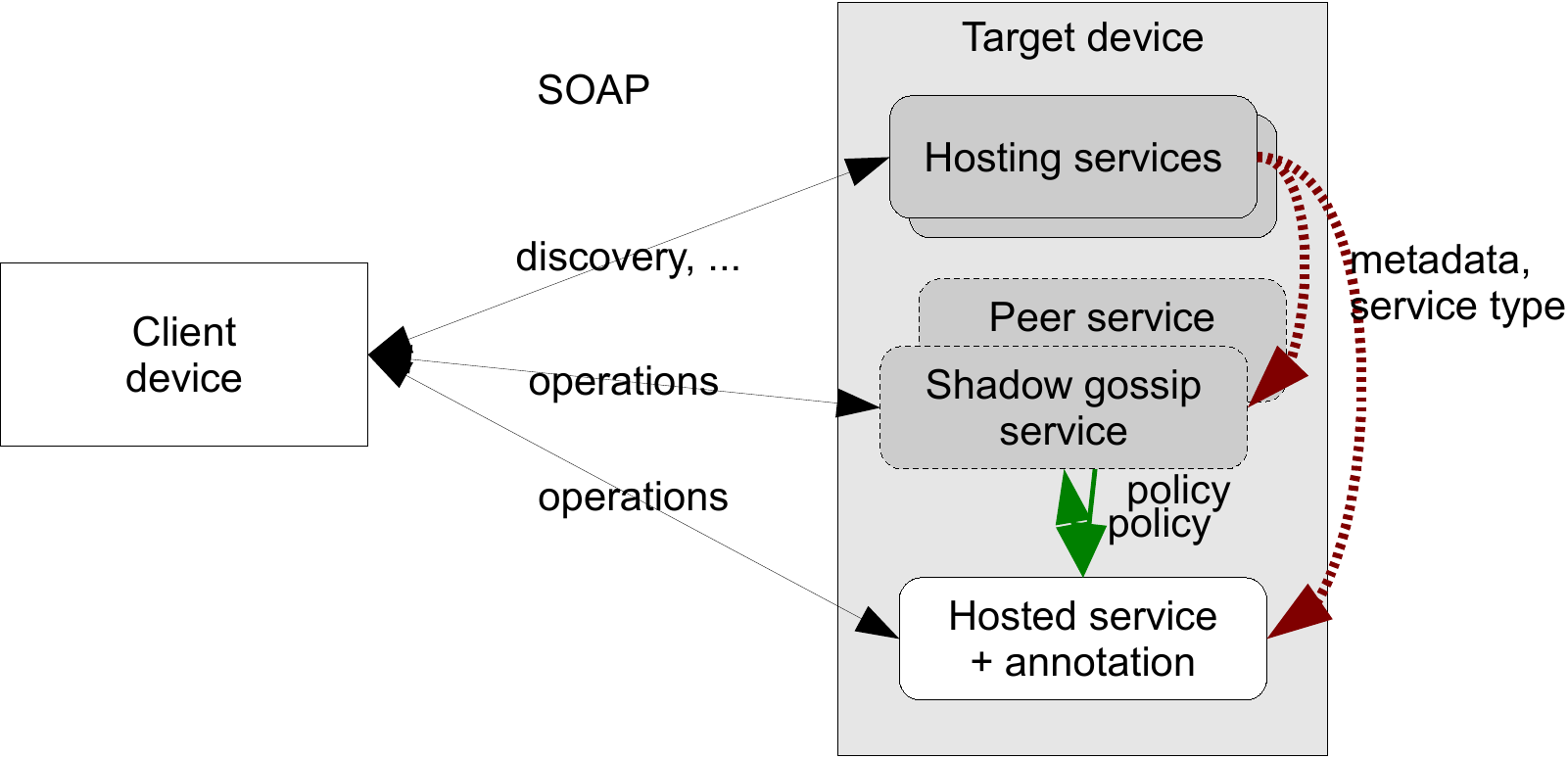}
\caption{Overview of WS-Gossip architecture.}
\label{fig:service}
\end{figure}

\begin{figure}[htbp]
\centering
\includegraphics[width=.8\textwidth]{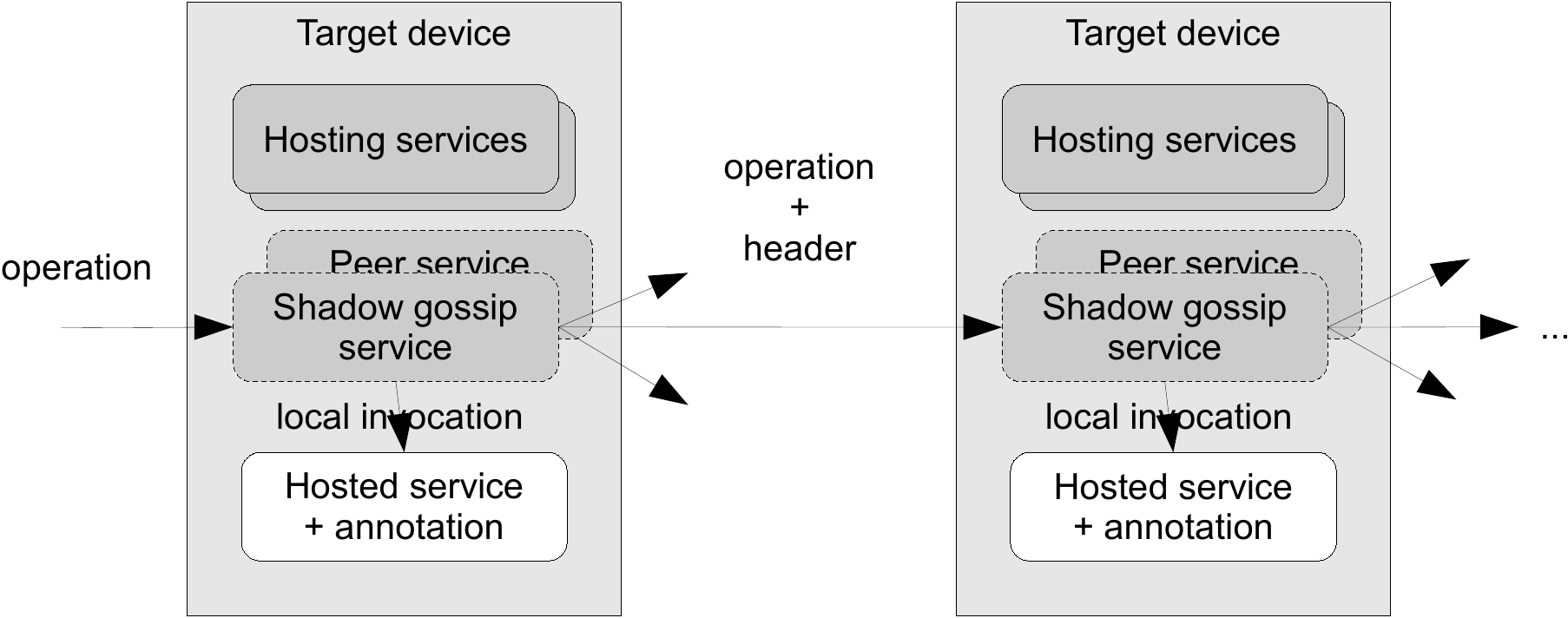}
\caption{Initialization of Gossip dissemination.}
\label{fig:gossip}
\end{figure}

Our proposal to address the scalability and reliability challenges of large DPWS deployments is to use a gossip-based dissemination protocol\,\cite{sigops-gossip-issue}. Gossiping is inherently scalable, as it spreads the load across its participants. Moreover, it is also inherently robust, tolerating message loss and participant crashes. This should have the increased advantage of allowing the usage of SOAP-over-UDP even if reliable delivery is desired, which is much less resource consuming than a full fledged HTTP binding over TCP. Moreover, by assuming the Web Services infrastructure, we take advantage of each gossiped unit of data being a SOAP envelope, of the self-documenting nature of services through WSDL, and of further standards such as WS-Addressing and WS-Policy.

Providing comprehensive support for gossip-based information dissemination in Web Services, in a way that integrates with existing DPWS deployments, thus reduces to the following challenges:
\begin{itemize}
 \item How to enable the usage of gossip by devices and clients, while at the same time minimizing the impact on producers and consumers of events, namely, regarding required middleware?
 \item How to support different peer discovery strategies, fit for different system scales and dynamics?
\end{itemize}
We address these challenges with a set of specifications of service port types, SOAP headers, and policy assertions that can be used to compose a variety of solutions. The general architecture of the proposed gossip service is outlined in Figure~\ref{fig:service} and works as follows. A manufacturer that intends to provide gossip dissemination in its devices can use a DPWS stack with gossiping support and annotate every service  supporting gossip using WS-Policy\,\cite{ws_p} assertions. As a consequence, a shadow gossip service is created for each service where gossip is enabled. Moreover a peer service can be setup to provide an entry point to the set of target peers. Multiple shadow gossip services can be attached to the same peer service, if they have the same set of targets.

Both the original hosted service and its shadow gossip service are advertised to clients that can use each of them independently. A gossip-aware client can examine policy annotations in both these services and determine their relationship. A client may still address the original hosted service, thus maintaining compatibility with existing clients that are unaware of gossiping.

Assume for now a \textit{one-way} or \textit{notification} operation (i.e. input or output only)\,\cite{wsdl_11,wsdl_12} and \textit{push} gossip\,\cite{892324}. Gossiping is started when a client sends a SOAP message to a port in the shadow gossip service. Note that this service exports the same port type as the original hosted service, which means that a legacy client can still be used, simply by invoking the endpoint of the new service. Upon reception of this message, it is inspected to determine if it contains a WS-Gossip header. If not, default gossiping parameters are obtained, including gossip variant, fanout, peer scope (according to WS-Discovery), and target binding (HTTP or UDP). Gossiping is then initiated by adding the gossip information to the message header and relaying it to a number of peers and to the local hosted service, as outlined in Figure~\ref{fig:gossip}.
When a gossip message is received, the gossiping interaction is continued by decrementing its hop count and by forwarding it to the selected peers. Note that such a message can be generated by a target device, as depicted in Figure~\ref{fig:gossip}, but it can also be generated directly by a gossiping-aware client. This allows a client to initiate gossiping in a custom scope or with custom parameters to achieve its own reliability and scalability trade-offs.

The remainder of this section explains in detail the information contained in SOAP headers, how the shadow service supports multiple gossiping and SOAP operation styles, and how the target set of peers is discovered and managed.

\subsection{Header information}

As previously stated, the unit of information being gossiped is the SOAP envelope. Messages in a gossip interaction contain an entry in the SOAP header section of the SOAP envelope describing how to relay such messages. These are initialized by the initiator device, either within a shadow service or by a gossip-aware client. Moreover, there is also the assumption of WS-Addressing\,\cite{ws_a} providing a unique identifier for each message and support for asynchronous replies. Briefly, it contains the following information:
\begin{description}
\item [Scope/Type]As defined by WS-Discovery, this field implicitly describes the set of targets. Devices can be configured to relay messages only within a specific scope and type.
\item [Fanout]The number of peers to target in each interaction.
\item [Hops]The remaining number of hops. This must be decremented by each device that relays the message. The message is discarded when it reaches zero.
\item [IdTTL]The time that each device should buffer the message identifier for duplicate detection. If this is set to zero, the protocol degenerates to the \textit{balls-and-bins} variant\,\cite{ballsandbins}.
\item [DataTTL]The time that each device should buffer the message itself for retransmission in lazy gossip variants. If this is set to zero, the protocol will never issue advertisements and will always use an \textit{eager} variant.
\item [Filter]An optional item, specifying a rule to filter replies, which must be specified using XSLT. Valid rules are configured by the deployer and advertised as policies by the shadow service.
\end{description}

\subsection{Operation styles}

SOAP and WSDL support several operation styles\,\cite{soap,wsdl_11,wsdl_12}. Besides a typical client-server interaction (i.e. \textit{request-response}), it is also possible to have input-only operations (i.e. \textit{one-way}), output-only operations (i.e. \textit{notification}), and call-back operations (i.e. \textit{solicit-response}). It is also possible that a \textit{two-way} operation leads to multiple replies. These different operation styles allow WS-Gossip to support different gossip variants in addition to the previously described eager \textit{push-style}, such as the \textit{lazy} and the \textit{pull} variants.

Gossiping in \textit{one-way} and \textit{notification} operations is handled as described previously: Upon reception of a message, it is propagated and no reply is expected. In \textit{request-reply} and \textit{solicit-response} operation, the message is propagated and then all replies received are propagated back to the initiator. This requires the initiator's address to be stored alongside with the message identifier used for duplicate detection during the specified \textbf{IdTTL}. Consider the following example: A \textit{request-response} to query available disk space of servers in a data center. A client invokes the operation on the shadow service, which eventually reaches all targets. All responses then travel back along the same tree implicitly created by the request message and will eventually reach the initiator.

An alternative is to make use of a filter. This can omit or aggregate replies according to a rule specified when gossip is initiated. 
Consider the following example: The same \textit{request-response} operation is used to determine which server has the most available disk space in a data center. This requires that upon deployment, devices are configured to support the maximum filter on the disk space query operation. A client invokes the operation on the shadow service, which eventually reaches all targets. Responses then travel back along the same tree implicitly created by request message, but they are buffered and filtered such that only the maximum discovered downstream is returned by each peer. Each peer's reply is sent as soon as all its targets have replied, with a value or with a fault, or when a timeout expires.

\subsection{Gossip styles}

In addition to eager \textit{push-style} gossip described so far, \textit{lazy} and \textit{pull} variants are supported as follows. Besides offering the same port type as the hosted service, the shadow gossip service provides a gossip port with the following operations:
\begin{description}
\item [Push]Alternative to directly using the interface. This allows a set of messages to be submitted in a single interaction.
\item [PushIds]Informs the target that a number of messages are locally available. These should then be requested using \textbf{Fetch}.
\item [Pull]Returns currently buffered messages during a time interval specified as a parameter.
\item [PullIds]Variant of the previous operation, which requests identifiers instead of the actual messages. These can then be requested using \textbf{Fetch}.
\item [Fetch]Returns currently buffered messages, as specified by a list of identifiers provided as a parameter.
\end{description}
Gossip variants can be achieved through the composition of the previous operations. Namely, \textit{lazy push} is obtained by using \textbf{PushIds} instead of \textbf{Push} and then waiting for \textbf{Fetch} to be used later on selected identifiers. \textit{Eager pull} is obtained by periodically invoking \textbf{Pull}. Finally, \textit{lazy pull} is obtained by periodically invoking \textbf{PullIds} and then using \textbf{Fetch} on the resulting identifiers that are unknown.

The gossip variant chosen for each operation depends on configuration by the service deployer. In particular, the optimum configuration for \textit{push} gossip is to use the \textit{eager} variant for early rounds and then \textit{lazy}. For \textit{pull} gossip, the \textit{lazy} variant is interesting for very large payloads. The combination of both \textit{push} and \textit{pull} is known to ensure rapid and robust dissemination of information\,\cite{892324,312207}.

\subsection{Peer service}

\begin{figure}[htbp]
\centering
\includegraphics[width=.75\textwidth]{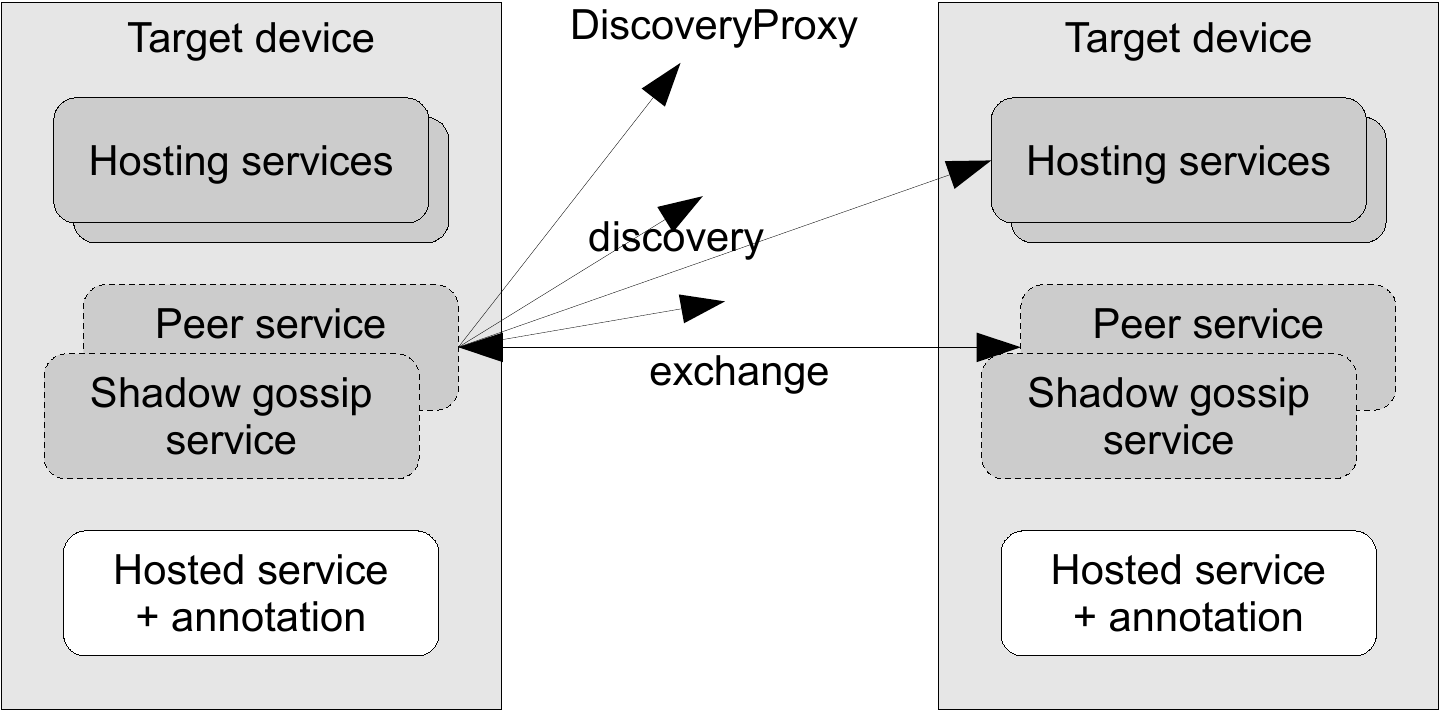}
\caption{Overview of peer management.}
\label{fig:memb}
\end{figure}

By default, WS-Gossip does not need an explicit peer management service. Instead, each gossip interaction can be configured with a scope or a service type that is then used to discover the full set of reachable peers through WS-Discovery. This is most useful in scenarios where a discovery proxy device exists, since a set of peers can be obtained efficiently by querying the proxy. This leads to a configuration with centralized peer information while information dissemination is distributed, which is adequate for scenarios with low churn and relatively high messaging rate.

If a proxy is not available, the usage of the Ad-Hoc mode of WS-Discovery would lead to a large number of multicast messages that would most likely defeat the purpose of gossip. Instead, our proposal allows that peers discovered to be cached locally and exchanged with other peers to implicitly create an overlay network using the Newscast protocol\,\cite{newscast03}.
The structure of the stored peer information comprises a list where each service instance is represented by an entry that contains the following elements:
\begin{description}
\item [Address]Corresponds to the service endpoint address.
\item [Type]Identifies the type of the service.
\item [DeviceId]Identifies the device where the service is hosted. This field is not applicable to services that are not associated with any device.
\item [Heartbeat]Counter that is incremented as messages, such as the invocation of the \textbf{Exchange} operation, are issued by other peer.
\end{description}
This information is exchanged among different devices and also updated through the examination of WS-Dis\-cov\-er\-y multicast messages issued by target services entering or leaving the network.
Periodically, if the instance has not received a request for exchanging its membership information during a certain time frame, it selects another instance of the peer service to which it sends such a request containing the list of the known endpoints. Upon reception of such a message, the contacted instance returns to the requester its own list of known endpoints, and merges it with the received one.

The heartbeat counter of a service instance that never sends a new message, or eventually sends but without reaching a peer service instance, stays unchanged, implying that it will move towards the end of the membership list as the counter of other services is being updated and new services are discovered. That service instance will eventually be discarded when the cache of the Peer Service reaches the configured maximum size.

\section{Performance Evaluation}
\label{sec:impl}
To evaluate the performance of the proposed approach, WS-Gossip was implemented and compared to the WS-Eventing implementation provided by
version 2 beta 3a of Java Multi Edition DPWS Stack (JMEDS), part of the Web
Services for Devices (WS4D) project\,\cite{ws4d_url}. The components of
WS-Gossip, both the shadow gossip service and the peer service were implemented
as regular hosted services, being able to coexist in the same device. For
testing, we have also implemented a simple service that exports a simple \textit{one-way}
operation to set the value of a float variable, mimicking the propagation of temperature values. By minimizing the payload, we
highlight the overhead of the protocol.

\subsection{Experimental setting}

Experimental evaluation is done using the Minha middleware test platform\,\cite{Carvalho:2011:EED:2093185.2093188,minha_url}, which virtualizes multiple devices within a single JVM while simulating the performance characteristics of a real system.
It also allowed us to inject network faults to better assess the reliability of WS-Gossip.

Each test corresponds to the simulation of the runtime of a given number of devices collocated in the same LAN, in a single host with the following configuration: 64-bit Ubuntu Server 10.04.4 Linux, two 12-core AMD Opteron\textsuperscript{TM} Processor 6172, 2.1GHz, 128 GB RAM, 64-bit Sun Microsystems Java SE 1.6.0\_26.

The evaluation consists in executing a periodic event dissemination, for the mentioned scenarios, where a new value is propagated from a single producer device to a given number of consumer devices.
A centralized managing device was used to control peer management\footnote{Discovery proxy is not yet implemented in JMEDS 2.0 beta 3a. Instead we used a custom registry service.} and the execution of the test.

The following scenarios were analyzed:
\begin{description}
\item [WS-Eventing]A publish/subscribe communication protocol was selected as it is one of the most used event dissemination patterns. Hence, the WS-Eventing standard, as provided by JMEDS, was evaluated using HTTP/TCP communication.
\item [WS-Gossip]The \textit{push} variant of WS-Gossip was selected to be evaluated in conjunction with SOAP-over-UDP.
Two different scenarios were evaluated for WS-Gossip in terms of communication errors to compare the achieved reliability and latency degradation.
Each of these scenarios designation is then suffixed with (0\% Loss), when there are no message losses, and with (10\% Loss), when 10\% of communication losses are introduced by the Minha simulator.
\end{description}

The execution procedure of each test comprised the following steps:
\begin{enumerate}
 \item The manager and the producer devices are started.
 \item The consumer devices are then started. In WS-Eventing,
they subscribe with the producer as soon as they are started. In WS-Gossip, the manager, informs each consumer of its neighbors according with the configured fanout value, so they can convey new messages to them.
For  both scenarios, the manager verifies if all the devices started correctly before signaling the producer to start the dissemination.
 \item The producer begins disseminating events periodically, which are propagated across the network.
 \item The producer terminates and notifies the manager.
 \item The manager informs, sequentially, all the devices about the file they should write their run statistics to.
\end{enumerate}
The tests for each scenario consisted in 5 runs for each given number of devices, where 120 events were periodically emitted with an interval of 5 seconds.

The interval between the initial emission of a message and its reception by a consumer was measured in nanoseconds since Minha enables the execution of all the intervening devices inside a single JVM on a single host.
The sampling of the instant of emission was performed right before the producer sends a message, and the reception time measurement was done in the first operation of the method invoked to deal with a new message at a consumer.

In WS-Gossip, the used values for the fanout parameter were computed according to\,\cite{amk-from-epidemics-to-dc}, taking into account the number of devices, as well as an expected error rate (e) of 5\% and a delivery assurance (p) of 99\%, ranging from a value of 8 for 10 devices to 11 for 250 devices.
In these very same scenarios, the publisher is randomly selected from all the nodes, contrarily to the WS-Eventing scenario where the publisher is the first device.

\subsection{Results and discussion}

\begin{figure}[htbp]
 \centering
 \begin{tikzpicture}
  \begin{axis}[xlabel=devices,ylabel=ms,ymax=130, legend columns=1, legend style={
 	at={(1.03,0.48)},
	anchor=west}]
   \addplot table[header=false,x index=0,y index=1]{pushudp-0.dat};
   \addplot table[header=false,x index=0,y index=1]{pushudp-10.dat};
   \addplot table[header=false,x index=0,y index=1]{notif.dat};
   \legend{WS-Gossip(0\% Loss),WS-Gossip(10\% Loss),WS-Eventing}
  \end{axis}
 \end{tikzpicture}
 \caption{WS-Eventing vs. WS-Gossip (latency).}
 \label{fig:wse_wsg}
 \end{figure}
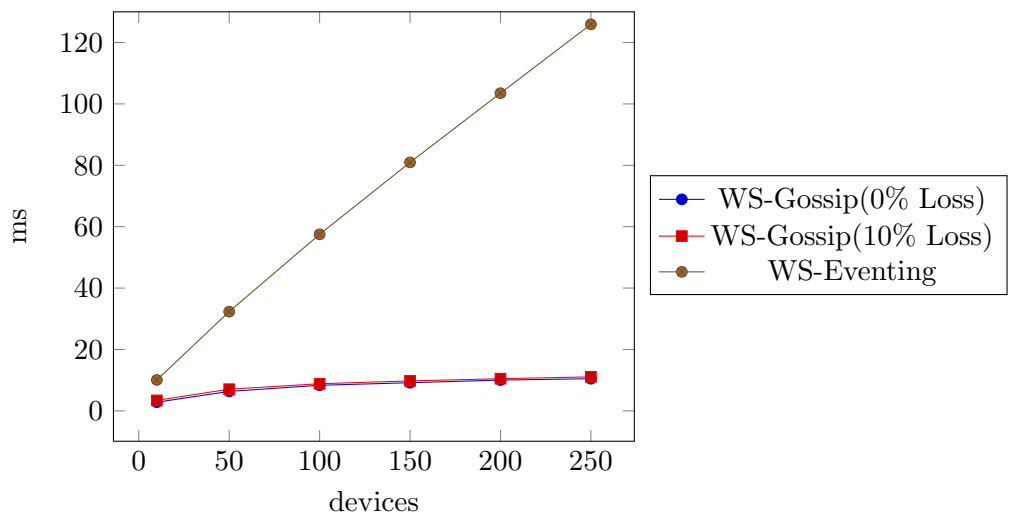
 
\begin{figure}[htbp]
 \centering
 \begin{tikzpicture}
  \begin{axis}[xlabel=devices,ylabel=hops,ymin=0, legend columns=1, legend style={
 	at={(1.03,0.48)},
	anchor=west}]
   \addplot table[header=false,x index=0,y index=3]{pushudp-0.dat};
   \addplot table[header=false,x index=0,y index=3]{pushudp-10.dat};
   \legend{WS-Gossip(0\% Loss),WS-Gossip(10\% Loss)}
  \end{axis}
 \end{tikzpicture}
 \caption{Average hops to delivery in WS-Gossip.}
 \label{fig:hops}
\end{figure}
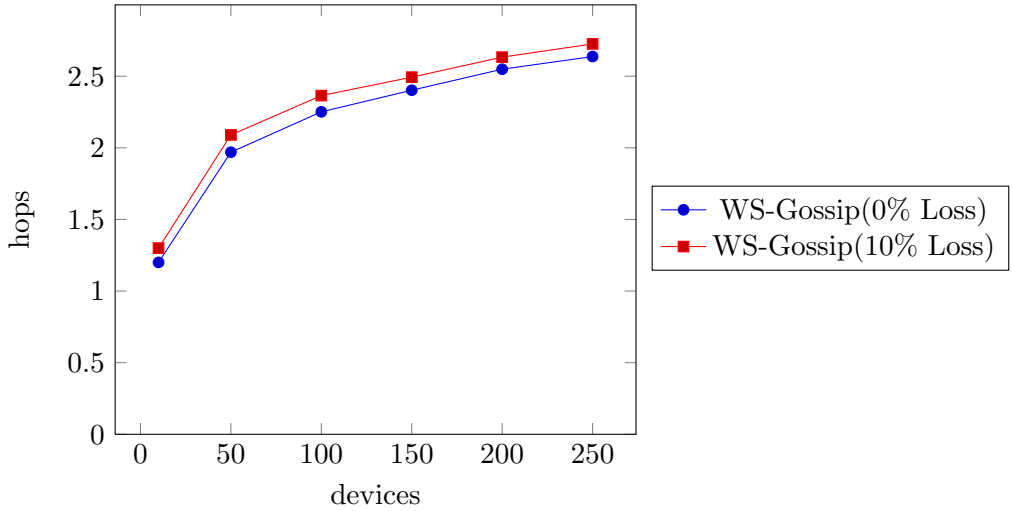

Results presented in Figures~\ref{fig:wse_wsg} and \ref{fig:hops} are the average of all 5 runs for each scenario.
For latency measurements, the first and the last 10 iterations were discarded in order to minimize the effect of Java JIT compilation, although it also masks the delay of TCP connection establishment in WS-Eventing.
 
In Figure\,\ref{fig:wse_wsg}, the message delivery latency of the WS-Eventing grows linearly with the number of targets, from 10 to 126 milliseconds, whereas that of WS-Gossip(0\% Loss) is very small and grows very slowly, between 2.8 to 10.5 milliseconds.
This can be justified by scattering the load of propagating a message throughout an entire network, by the devices on that network, instead of overloading a single device, such as the publisher in WS-Eventing.
The message delivery latency of WS-Gossip(10\% Loss) is very close to that of WS-Gossip(0\% Loss), suffering a small increase of around 0.1 milliseconds.
 
Figure\,\ref{fig:hops} presents the logarithmic growth of the average number of hops a message goes through from emission to reception in WS-Gossip, between 1.2 to 2.64 hops, confirming that the gossip protocol scales logarithmically with system size. This figure also shows that the introduction of communication losses has little effect on the number of hops a message goes through, with an average increase of around 0.6 hops in WS-Gossip(10\% Loss) compared to the baseline scenario, where no messages are lost in the network.
 
Message delivery rate is not presented graphically since it is 100\% both in WS-Eventing and in WS-Gossip(0\% Loss) and it is always greater than 99.9\% in WS-Gossip(10\% Loss), and most frequently 100\%, even with a rate of communication losses that corresponds to the double of the expected 5\%.

To conclude the analysis of the results, considering an environment with \textit{n} devices, where the WS-Eventing producer will always have to send \textit{n} messages for each event, whereas gossip peers will send a number of messages equal to its fanout \textit{f}, thus spreading the load throughout the network which results in savings in the consumption of resources by the producer for cases where \textit{n} $ > $ \textit{f}.

\section{Related Work}
\label{sec:related}
Service Oriented Computing has proven to be a very adaptable programming
paradigm, even to an environment with scarce resources such as Wireless Sensor
Networks\,\cite{Mohamed:2011p18703}, where Service Oriented Middleware
should fulfill some stringent requirements. DPWS, as a standard specially
targeted to enable the usage of Web Services by resource constrained devices,
already provides several of the required features, such as dynamic, adaptive and
auto-configurable architectures. Therefore, DPWS is a key standard for the
implementation of the Internet of Things paradigm, and it has, for instance,
become the enabler of the Smart Grid\,\cite{Karnouskos:2012p18459}, by allowing
the interaction of entities with largely heterogeneous processing power, namely,
smart meters and utilities backend energy management systems, albeit indirectly
through an hierarchical architecture. This interaction allows a better
monitoring of all the online assets and a better control of power generation, as
it can be adjusted to better suit the demand.

Regarding communications among devices in a LAN, an extension for the usage of
UDP Multicast with WS-Eventing was proposed\,\cite{Gregorczyk:2011p17697}, which
could help reduce the amount of traffic in scenarios where a single publisher
must inform various subscribers on the occurrence of periodic events. However,
the assurance of reliable delivery of events using a positive acknowledgment
system would cause an acknowledgment explosion in the publisher. And even in the
case of well-known periodic events, where the usage of notification
retransmission requests would be suitable, it could lead to a similar scenario
if various subscribers do not receive the same event, since each will trigger a
retransmission request which will ultimately accumulate on the publisher's side.

As WS-Coordination and other related standards provide transaction support to
Web Services as a fault tolerance mechanism, there is still room for
improvement, not only in terms of the actual transaction
modeling\,\cite{Papazoglou:2007p17304}, but also in terms of failure recovery.
For instance, a flexible compensation mechanism to perform backward failure
recovery has been added to WS-BusinessActivity\,\cite{Liu:2008p18821} in order
to improve the dependability of long-running business transactions, by allowing
participants to select from alternative compensation operations for each Web
Service operation instead of a single compensation operation as defined in the
standard.

WS-Membership\,\cite{Vogels:2003p1034} proposed a framework that provides cooperating Web Services and
activity monitors with a unified approach for tracking registered Web Services
and for supplying membership updates to monitors using gossip-style
communication, hence, promoting an highly robust and asynchronous membership
information propagation mechanism with good reliability and scalability
capabilities. However, it was not standardized and seems to have ceased to exist
as little information can be found on the internet.
Albeit the disadvantages of using epidemic failure detection, like inefficiency
when the size of messages grows proportionally with the number of participants,
and bad behavior with massive concurrent participant failures, the detection of
failed Member Services is very accurate.

\section{Conclusion}
\label{sec:concl}
Information dissemination in the context of service oriented architectures involving large numbers of connected devices poses a set of challenges that are not adequately met with traditional approaches.
To address these challenges, we propose the usage of gossiping at an architectural level instead of either relegating the information dissemination problem to black box middleware or coping with the limitations of heavyweight coordination protocols and their assumptions of buffering and transactional logs for reliability.

Gossiping has several advantages in this context, as a variety of gossip protocols can be achieved with minimal complexity while providing strong guarantees of reliable and atomic dissemination.
Moreover, our proposal supports such variations with a simple service interface and a set of possible interactions, which include both one-to-many and many-to-many dissemination, as well as many-to-one aggregation queries.
In contrast to previous approaches\,\cite{762485}, our proposal integrates seamlessly in a Device Profile for Web Services (DPWS) environment, being compatible with existing devices.
By implementing the proposed architecture on the Web Services for Devices (WS4D) Java Multi Edition DPWS Stack (JMEDS), we show that the performance of a one-to-many operation using gossip improves on bare SOAP-over-UDP, included in DPWS, both on latency and fault tolerance, while offering additional flexibility and resilience, largely surpassing the performance of WS-Eventing, as provided by JMEDS, in terms of latency.

For future work, we intend to evaluate other gossip variants provided by our framework and the effect of churn and multiple event sources in the network.

\section{Availability of Code}

The source code developed and used for the performance evaluation comprised in this paper is available as open source, allowing the experiments to be reproduced.
In detail, our implementation of WS-Gossip on the Web Services for Devices (WS4D) Java Multi Edition DPWS Stack (JMEDS) is available at \url{https://github.com/filipecampos/ws_gossip}.
The code used in the WS-Eventing scenarios, as well as for setting up and controlling all the experiments, is available at \url{https://github.com/filipecampos/ws_gossip_tests}.

\section*{Acknowledgments}
\small{
This work has been partially supported by the Portuguese National Science Foundation FCT - Fundação da Ciência e Tecnologia, through grant SFRH/BD/66242/2009.
}

\bibliographystyle{abbrv}
\bibliography{article}

\end{document}